\documentclass[aps,prl,numerical,reprint,superscriptaddress,showpacs]{revtex4-1}
\usepackage{color}
\usepackage[version=3]{mhchem} 
\usepackage{gensymb}
\usepackage{amssymb}
\usepackage{upgreek}
\usepackage{graphicx}
\usepackage{amsmath}
\usepackage{xr}

\begin{document}

\title{Network patterns in exponentially growing 2D biofilms}

\author{Cameron Zachreson}
\affiliation{School of Mathematical and Physical Sciences, University of Technology Sydney, Ultimo, NSW, 2007, Australia}

\author{Xinhui Yap}
\affiliation{The ithree institute, University of Technology Sydney, Ultimo, NSW, 2007, Australia}

\author{Erin S. Gloag}
\affiliation{The ithree institute, University of Technology Sydney, Ultimo, NSW, 2007, Australia}
\affiliation{Center for Microbial Interface Biology, Ohio State Univeristy, 460 W. 12th Ave., Columbus, OH 43210}

\author{Raz Shimoni}
\affiliation{The ithree institute, University of Technology Sydney, Ultimo, NSW, 2007, Australia}

\author{Cynthia B. Whitchurch}
\affiliation{The ithree institute, University of Technology Sydney, Ultimo, NSW, 2007, Australia}

\author{Milos Toth}
\affiliation{School of Mathematical and Physical Sciences, University of Technology Sydney, Ultimo, NSW, 2007, Australia}

\begin{abstract}
Anisotropic collective patterns occur frequently in the morphogenesis of 2D biofilms. These patterns are often attributed to growth regulation mechanisms and differentiation based on gradients of diffusing nutrients and signalling molecules. Here, we employ a model of bacterial growth dynamics to show that even in the absence of growth regulation or differentiation, confinement by an enclosing medium such as agar can itself lead to stable pattern formation over time scales that are employed in experiments. The underlying mechanism relies on path formation through physical deformation of the enclosing environment. 

\end{abstract}

\date{\today}

\maketitle

\begin{figure}
\centering
{\includegraphics[width=0.42\textwidth]{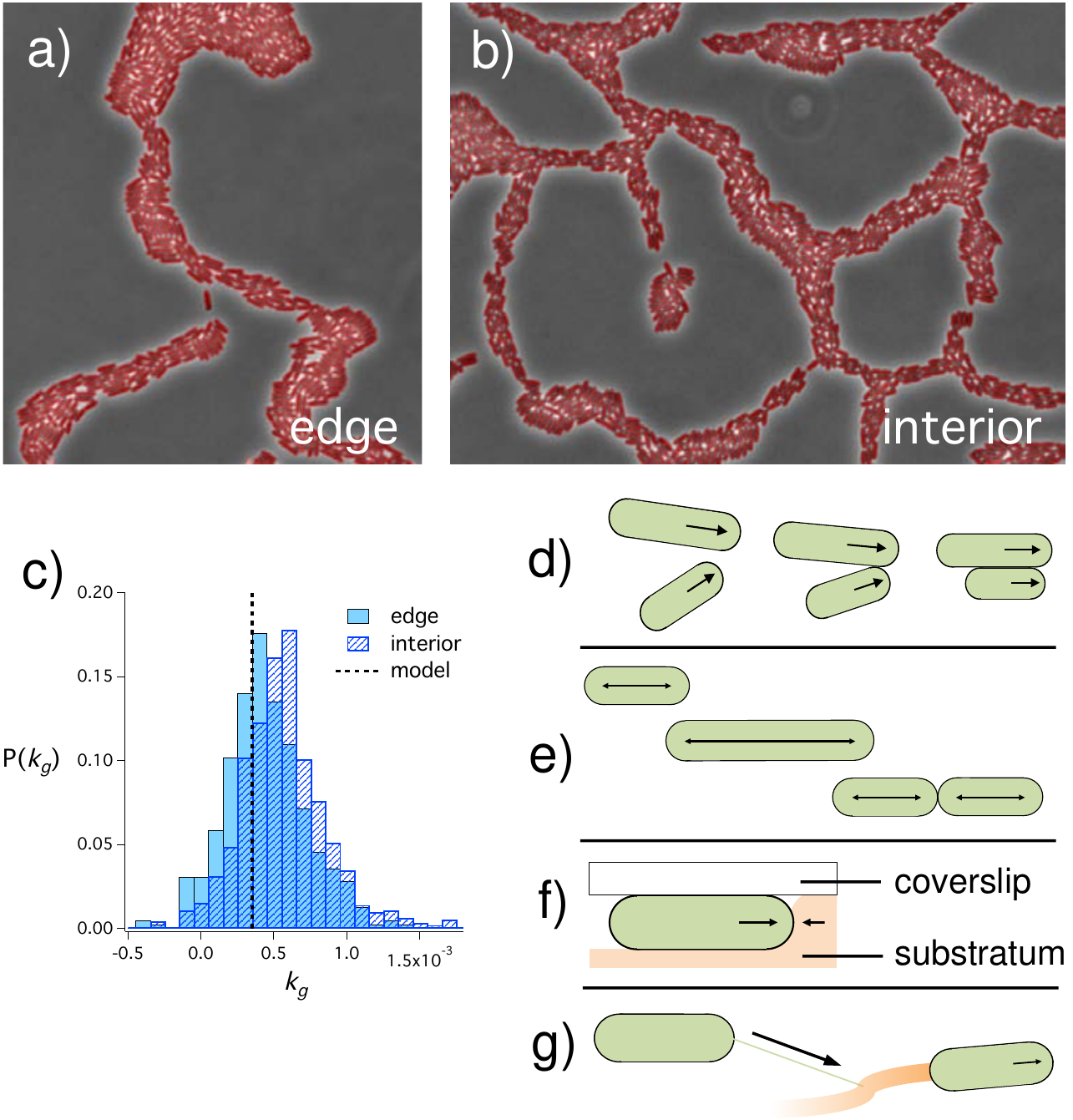}}
\caption{(a, b) Phase-contrast images of the edge and interior (respectively) of an expanding {\it P. aeruginosa} interstitial biofilm \cite{gloag2013self}. Each image is frame 500 from a total of 1000 used to make Supplemental Movies S1 and S2 \cite{Supp}. Binary masks of the cells produced by the image segmentation algorithm are shown in red. (c) Histograms of the probability $P$ for a cell to be growing at rate $k_g$, as calculated from the movies corresponding to the colony edge (a, solid bars in c) and interior (b, dashed bars in c). The dashed line represents the growth rate used in our simulations. (d-g) Schematic illustrations of the processes comprising our model of bacterial behavior.}
\label{Fig1}
\end{figure}

In surface-associated bacterial colonies, growth regulation caused by a diffusion-limited nutrient supply has been suggested as a primary mechanism of pattern formation, resulting in branched, locally anisotropic morphologies \cite{ben1994generic,farrell2013mechanically}. However, diffusion-limited growth does not tell the whole story. Mechanical processes are known to play essential roles in morphogenesis \cite{howard2011turing}, and we show here that deformation of the enclosing environment by moving bacteria can lead to the emergence of persistent, stable, anisotropic network patterns such as those in Fig.~\ref{Fig1}(a,b) even under conditions of exponential colony growth. Furthermore, we present strong evidence that in {\it Pseudomonas aeruginosa}, well-studied trail following mechanisms involving the deposition of extracellular polymeric substances (EPSs, or ``slime") cannot produce these patterns in the absence of such mechanical effects.

Interstitial biofilms form when bacteria are introduced to the space between two apposed surfaces and are constrained to a quasi-2D arrangement. If one surface is a hydrogel such as agar or gellan gum \cite{gloag2013self,semmler1999re}, the bacteria have access to nutrients diffusing through this medium \cite{schantz1962diffusion}. Here we consider such a situation, where nutrient depletion is negligible and the bacterial growth rate is not retarded in the colony interior. This assertion is supported by histograms of individual growth rates we obtained from of the edge and interior  of an expanding {\it P. aerugionsa} interstitial biofilm [Fig.~\ref{Fig1}(a,b), Supplemental Movies S1, S2 \cite{Supp} respectively]. The histograms show that growth rates measured in the colony interior and the colony edge are approximately equivalent [Fig.~\ref{Fig1}(c)], and nutrient depletion therefore cannot be used to explain the network morphology seen in Fig.~\ref{Fig1}(a,b), characterized by dense trails of bacteria and voids of zero density.

\begin{figure}
\centering
{\includegraphics[width=0.42\textwidth]{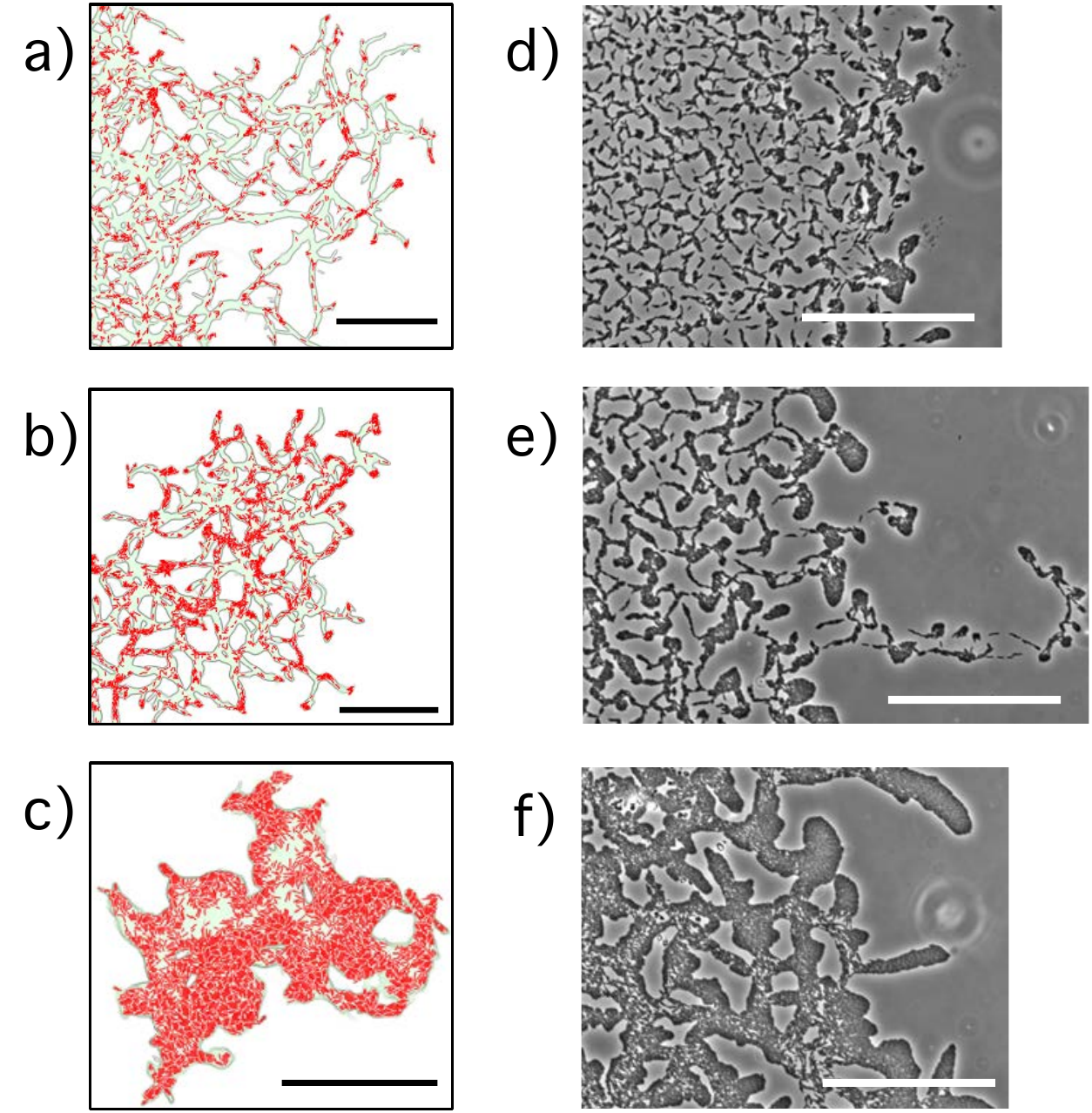}}
\caption{Snapshots of three biofilms simulated by the model with substratum stiffness coefficients of $\gamma = 0.25, 0.5, 1.0$ (a-c, Supplemental Movies S3-S5 \cite{Supp}, respectively), illustrating the cell positions (red), EPS trails (green) and furrows (gray outline). The simulations show qualitative agreement with experiments performed using hydrogel monomer (gellan gum) concentrations of $0.5\%, 0.8\%$, and $1.2\%$  (d, e, and f respectively). Scale bars represent 100 $w$ (a-c), or 100 $\mu$m (d-f). }
\label{FigES}
\end{figure}

Previously, Gloag et al. observed that these bacteria can deform the soft enclosing material through which they move, and confinement in the resulting `furrows' appears to contribute to network pattern formation \cite{gloag2013self}. However, {\it P. aeruginosa} and many other surface-motile bacteria \cite{pelicic2008type} move through the extension and retraction of type IV pili (T4P) \cite{skerker2001direct}, which have increased binding affinity to extracellular polymeric substances (EPS) secreted by the moving bacteria \cite{maier2015bacteria}, a phenomenon that has also been suggested to result in trail-following behavior \cite{Gelimson2016Multicellular,kranz2016trails}. To investigate the relative importance of these two  processes in the emergence of the observed network morphology, we carried out simulations of motile bacteria that interact with the hydrogel environment. The behavioral rules implemented in our simulations are illustrated in Fig.~\ref{Fig1}(d-g). Each simulation was initiated with a single motile, growing cell, and terminated after 12 cell division cycles ($t_{f} = 1.37\times 10^5~s$). Our results verify the importance of hydrogel properties in interstitial biofilm morphogenesis. By increasing the substratum stiffness in our model, colony morphology [Fig.~\ref{FigES}(a-c)] was altered in qualitative agreement with experimental results obtained by increasing hydrogel monomer concentration [Fig.~\ref{FigES}(d-f)]. [See \cite{Turnbull2014} for details on the interstitial biofilm culture techniques used to produce the data in Figs.~\ref{Fig1}(a,b) and \ref{FigES}(d-f).]


Bacterial behavior emerges from a complex interplay of many non-equilibrium and stochastic processes. Isolation of factors that are fundamental to morphogenesis mechanisms requires a model that describes the complexity of the real system, with an experimentally constrained parameter space. Here, we focus on the processes of cell growth and motility as the fundamental model ingredients. The essential properties of the system that cannot be estimated {\it a-priori} relate to bacterial movement dynamics and individual growth rates. We therefore designed our biophysical model so as to enable incorporation of experimentally determined time-scales into these processes.

The model used here is similar to one we reported previously \cite{zachreson2016emergent}, except that it includes cell growth and division.  We parameterized our model using experimental data that quantify single-cell dynamics \cite{skerker2001direct}, and our own measurements of individual growth rates [Fig.~\ref{Fig1}(c)] \cite{Supp}. 



Individual bacteria are simulated as self-propelled, capped rods of unit width $w$, that undergo repulsive collisions with one another [Fig.~\ref{Fig1}(d)], elongate at a constant rate $k_g$ and divide [Fig.~\ref{Fig1}(e)] at a critical length $l_{max} = 7~w$ into two cells each with length $l_{min} = 3~w~ \pm ~  \delta l $, where $\delta l$ represents asymmetric division and is randomly selected at the time of division from the uniform interval $\delta l \in [0, 1]~w$. Cell-cell collisions produce force and torque away from the contact point. It is also possible for cells to pull on one another with their pili because we model the cell motility mechanism explicitly \cite{zachreson2016emergent} by simulating the extension, binding, and retraction of T4P \cite{mattick2002type}. The equations for translational and rotational motion of rods and our repulsive interaction scheme are the same as those used in \cite{farrell2013mechanically,ghosh2015mechanically,zachreson2016emergent}. During T4P-mediated surface motility, {\it{P. aeruginosa}} cells spontaneously reverse polarity, this process is believed to be essential to chemotaxis phenomena, but polarity reversals occur stochastically even in the absence of attractant/repellant gradients \cite{oliveira2016single}. Here, stochastic polarity reversals were implemented using a `countdown' type algorithm where the time between reversal events is selected from a Gaussian distribution with a mean $\langle T_{rev} \rangle= 1000 s$ and a standard deviation $\sigma_{rev} = 200s$. 

Substratum deformation [Fig.~\ref{Fig1}(f)] is simulated by a surface potential $U$ that resists motion up its gradient. $U$ increases when a bacterium is present, representing a depression in the substratum. Three parameters control the mechanical properties of the substratum: the stiffness $\gamma$, and the deformation and restitution rates $k_U$ and $\beta_U$ of the potential $U$, respectively. The stiffness, $\gamma = dU/dC_s$ modulates the force $\vec{F}_{s}$ exerted by substratum deformation, where $C_s$ represents surface topography, such that $\vec{F}_{s} = -\gamma \nabla C_{s} $ is the force exerted due to deformation. The deformation rate per unit area, $k_s = k_U/\gamma$ and restitution rate $\beta_s = \beta_U/\gamma$, define the substratum dynamics for a spatial region of area $[\Delta x] ^2$ so that $\dot{C_s} = k_s[1 - C_s(t)/C_{max}][\Delta x] ^2 - \beta_s C_s(t)$, where $C_{max} = w$ and the accumulation term only applies if a bacterium is present. These rate constants $k_s$ and $\beta_s$ are inversely proportional to $\gamma$ and we hold their ratio fixed at $\beta_s / k_s = 0.1$ in order to ensure a consistent maximum trail depth. The substratum deformation and restitution give rise to path formation and degradation, because the force corresponding to the gradient of $U$ influences the motion of the bacteria. 

We simulate the function of secreted EPS in motility [Fig.~\ref{Fig1}(g)] with a trace left behind by bacteria that modulates the T4P binding probability: $P_s = KP_{max} + (1 - K)P_{min}$, where $K = C_p/[\Delta x]^2$ is the fraction of local area covered by binding sites for pili and $C_p$ is the local trace value that accumulates and degrades analogously to $C_s$ according to the EPS deposition and degradation rates ($k_{eps} = 0.1$ and  $\beta_{eps} = 5 \times 10^{-4}$, respectively).

A similar model of motility bias due to EPS secretion was published recently \cite{Gelimson2016Multicellular,kranz2016trails}. However, our model explicitly simulates the stochastic process of pilus binding and retraction \cite{zachreson2016emergent}, instead of making a mean-field approximation based on the assumption of a large number of T4P per cell. Based on available literature, such an assumption may not be valid for {\it{P. aeruginosa}} \cite{skerker2001direct}. This apparently subtle difference is significant because our findings suggest that the stochastic nature of T4P binding and retraction yields ineffective following of EPS trails that is not robust in the absence of furrowing. This is exemplified by Fig.~\ref{Fig2}(a), which illustrates a typical colony morphology in a case where physical interactions with the substratum are weak. In such a case trail networks could, in principle, emerge exclusively {\it{via}} the EPS stigmery mechanism but do not, making the EPS phenomenon alone unlikely to account for network formation in {\it P. aeruginosa}. A detailed discussion of this important result, including tests of model variants lacking substratum deformation, is provided in the Supplemental Information \cite{Supp}.

Because the furrowing process is governed by the substratum deformation rate $k_U$ and stiffness $\gamma$, these parameters control the path formation rate and degree to which cells are confined to the paths, respectively, making them critical to this trail formation mechanism. To investigate their roles, we simulated bacterial growth under various environmental conditions defined by $k_U$ and $\gamma$, and recorded the colony configuration (cell positions, lengths, and orientations) at $20~s$ intervals.

\begin{figure}[h]
\centering
{\includegraphics[width=0.42\textwidth]{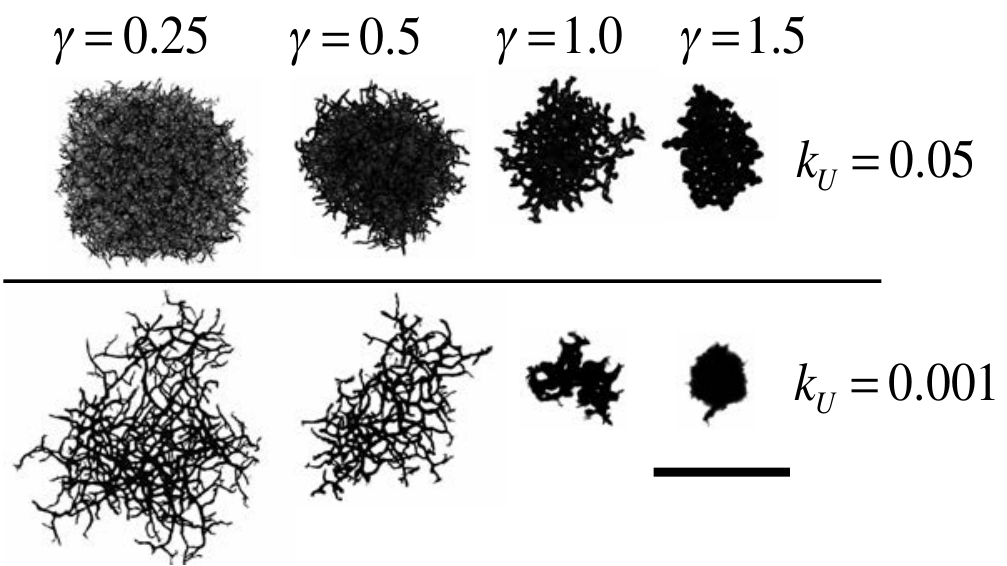}}
\caption{Plots of simulated furrow depth after 11 division cycles. Four different substratum stiffness coefficients ($\gamma$), and two different deformation rates ($k_{U}$) are represented. The scale bar represents 400 $w$.}
\label{Fig3}
\end{figure}

\begin{figure}[h]
\centering
{\includegraphics[width=0.42\textwidth]{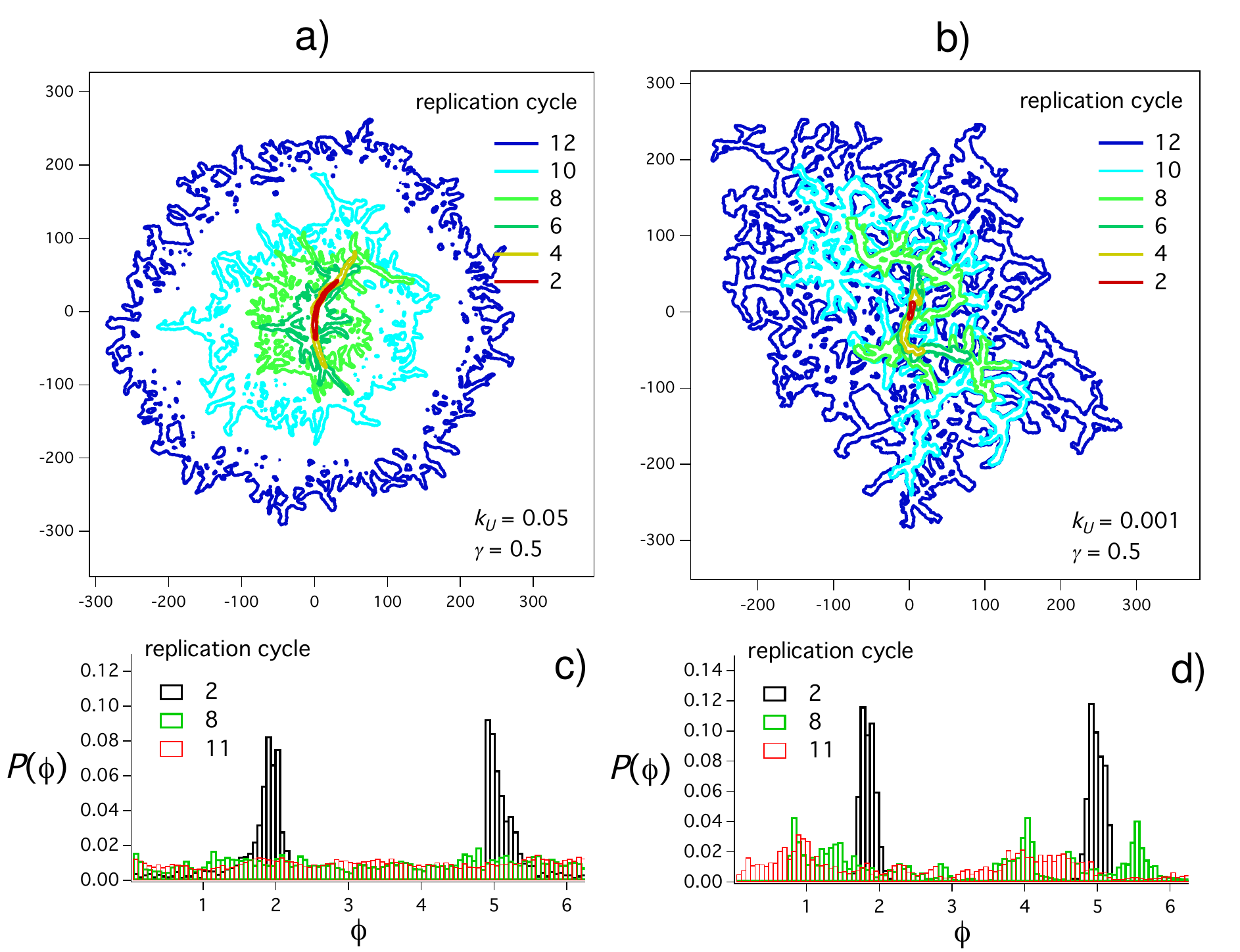}}
\caption{(a, b) Colony outline (defined as the area over which $C_s \neq 0$) plotted after a number of cell replication cycles on soft (a) and hard (b) agar (Supplemental Movies S6 and S4, respectively \cite{Supp}). (c, d) Histograms of polar coordinates $\phi$ of the constituent bacteria integrated over the indicated replication cycles for soft (c) and hard (d) agar.}
\label{Fig2}
\end{figure}

Three different morphological classes emerge as a function of the parameters defining the mechanical properties of the substratum. These consist of two qualitatively distinct isotropic states, and a locally anisotropic network similar to the one observed in experiments. Dense configurations become isotropic due to buckling \cite{boyer2011buckling} (Fig.\ref{Fig3}, $\gamma = 1.5$, $k_U = 0.001$), while dilute configurations correspond to isotropic diffusive behavior [Fig.~\ref{Fig3}, $\gamma = 0.25, 0.5 ~;~k_U = 0.05$]. 

Unregulated exponential growth eventually outpaces all other processes in the model and acts to homogenize the colony morphology as $t\rightarrow\infty$. However, on the time scales probed in typical experiments, the colony can rapidly become isotropic,  [Fig.~\ref{Fig2}(a)] or it can grow in a network [Fig.~\ref{Fig2}(b)], depending on the values of $k_U$ and $\gamma$. The changes in colony morphology during growth seen in Fig.~\ref{Fig2}(a,b) correlate with changes in the radial distribution of cells around the colony center of mass [Fig.~\ref{Fig2}(c,d)], which transitions from a multi-modal to a uniform distribution as the colony grows from a single cell moving back and forth  in the hydrogel to become isotropic. 

The entropy of such radial distributions as a function of length scale $r$ and distance from the colony edge $d_{edge}$ provides a metric of local anisotropy in the density distribution, a signature of the network morphology observed in experiment [Fig.~\ref{Fig1}(a,b), and Fig.~\ref{FigES}(d,e)]. Our anisotropy parameter $\Delta S_{\phi}(r, d_{edge})$ reflects the entropy of local distributions relative to that of the uniform distribution. $\Delta S_{\phi}$ takes a value of 1 for a 1D density distribution and approaches 0 for an isotropic (circular) distribution \cite{Supp}. 

\begin{figure}
\centering
{\includegraphics[width=0.45\textwidth]{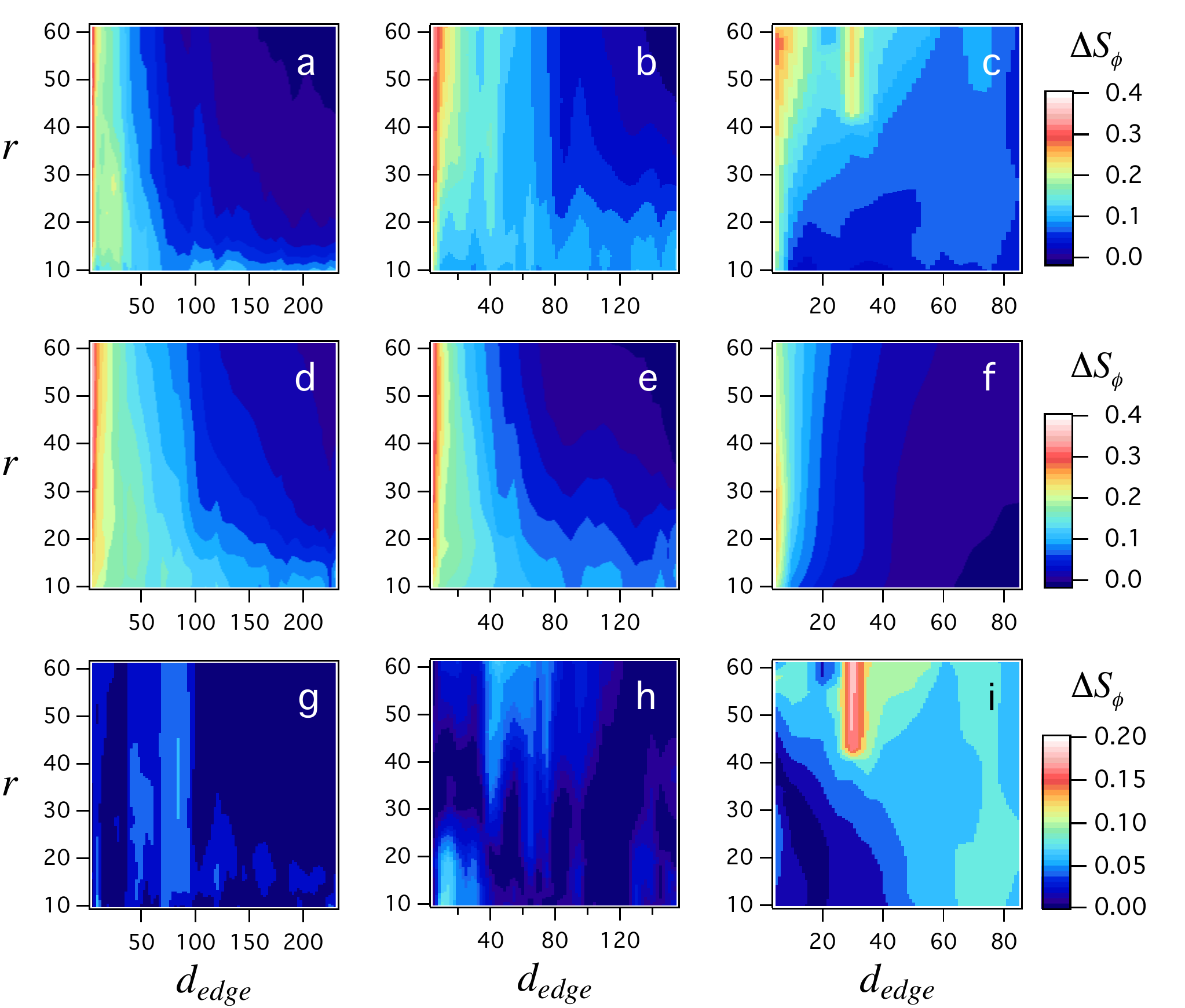}}
\caption{Intensity maps of $\Delta S_{\phi}$ as a function of length scale $r$ and distance from the colony edge $d_{edge}$. Measurements from images of bacterial biofilms grown on 0.5\%, 0.8\%, and 1.2\%  gellan gum substrata (a, b, and c respectively), are compared to data measured from simulations with increasing $\gamma$ values of 0.25, 0.5, and 1.0, (d, e, and f). Residuals comparing simulation results to experimental measurements are shown in g, h, and i ( $g = |a-d|$, $h = |b-e|$, and $i = |c - f|$ ). All distances are in units of average cell width $w$.}
\label{dS_exp_sim}
\end{figure}

Quantitative comparison of $\Delta S_{\phi}$ in the simulated and experimental morphologies depicted in Fig.~\ref{FigES} are provided in Fig. \ref{dS_exp_sim}. The results indicate good agreement in cases where network morphology is observed [Fig.~\ref{dS_exp_sim}(g, h)], with high degrees of anisotropy at all length scales near the colony edge, falling off to shorter length scales towards the interior. For the dense morphology, quantitative similarity is confined to shorter length scales, close to the colony edge (Fig.~\ref{dS_exp_sim}i). This discrepancy is likely due to greater cell numbers and colony area in experiment, as well as an overestimation of the substratum resistance value $\gamma$ in the corresponding simulation. Qualitatively, however, there is an inversion in the scaling of $\Delta S_{\phi}$ as a function of $r$ between the network and dense morphologies in both simulation and experiment. In the dense morphology, anisotropy increases at longer length scales, while in the network morphology it is highest at shorter length scales corresponding to narrow trails.

Due to mechanical constraint from the substratum, movement of the initial bacterium is either negligible or quasi-1D due to resistance to orientational and translational movement. As cell numbers increase, this 1D confinement facilitates collisions between cells that produce enough force to change cell orientations, causing branching. The dense configuration occurs when motility cannot outpace volume expansion due to population growth (Supplemental Movies S5 and S7 \cite{Supp}). The diffusive mode corresponds to the situation where collisions between small numbers of cells are sufficient to overcome the orientational resistance applied by the substratum. In this case, branching occurs readily in arbitrary locations within an existing trail and the system becomes isotropic as branches interlink (See Movie S6 \cite{Supp}). Network morphology occurs in the intermediate case, where collisions between larger numbers of cells are required for branching. The process can be observed in Supplemental Movie S4 \cite{Supp}. 

If EPS deposition is ignored and the attachment probability $P_s$ is uniform, qualitatively different trail-network patterns are possible [Fig.~\ref{SR2}(c,d)] that evolve from a randomly diffusing state due to the amplification of fluctuations in surface topography (Supplemental Movie S8 \cite{Supp}). These patterns are distinct from those formed by our simulations of {\it {P. aeruginosa}} (Fig.~\ref{dS_r_d}, \cite{Supp}). 

To conclude, we have simulated biofilm growth to explain how network morphogenesis in expanding bacterial colonies arises from an interplay between mechanical confinement and enhanced motility due to EPS deposition. These patterns cannot be explained by the EPS trail following mechanism alone, and emerge despite the tendency for exponentially growing colonies to expand isotropically. Moreover, their formation does not require chemotaxis, contact-based signaling, surface adhesion between cells, or EPS-mediated nematic alignment. 
\\\\
CZ would like to acknowledge Christian Wolff and Matthew Arnold for useful discussions. This work was funded by FEI Company and the Australian Research Council (DP140102721).


\newcommand{\beginsupplement}{%
        \setcounter{table}{0}
        \renewcommand{\thetable}{S\arabic{table}}%
        \setcounter{figure}{0}
        \renewcommand{\thefigure}{S\arabic{figure}}%
        \setcounter{page}{1}
        \renewcommand{\thepage}{S\arabic{page}} 
     }

\section*{Supporting Information for\\`Network patterns in exponentially growing 2D biofilms'}
\centerline {by Zachreson \it et al.}
\hspace{1cm}
\beginsupplement
\newcommand{\uvec}[1]{\boldsymbol{\hat{\textbf{#1}}}}
\subsection{Supplemental movies}

\begin{itemize}
\item Movie S1: The advancing edge of a {\it{Pseudomonas aeruginosa}} interstitial biofilm (2000s, 0.5 frame/s).
\item Movie S2: The interior of a {\it{Pseudomonas aeruginosa}} interstitial biofilm (2000s, 0.5 frame/s). 
\item Movie S3: Simulated interstitial biofilm showing bacteria in red, EPS trails in green and substratum deformation in gray-scale. $\gamma = 0.25$, $k_U = 0.001$, ($1.35 \times 10^5s$,  1000s/frame). 
\item Movie S4: Simulated interstitial biofilm showing bacteria in red, EPS trails in green and substratum deformation in gray-scale. $\gamma = 0.5$, $k_U = 0.001$, ($1.35 \times 10^5s$,  1000s/frame). 
\item Movie S5: Simulated interstitial biofilm showing bacteria in red, EPS trails in green and substratum deformation in gray-scale. $\gamma = 1.0$, $k_U = 0.001$, ($1.35 \times 10^5s$,  1000s/frame). 
\item Movie S6: Simulated interstitial biofilm showing bacteria in red, EPS trails in green and substratum deformation in gray-scale. $\gamma = 0.5$, $k_U = 0.05$, ($1.35 \times 10^5s$,  1000s/frame). 
\item Movie S7: Simulated interstitial biofilm showing bacteria in red, EPS trails in green and substratum deformation in gray-scale. $\gamma = 1.5$, $k_U = 0.001$, ($1.25 \times 10^5s$,  1000s/frame). 
\item Movie S8: Simulated interstitial biofilm with EPS trail following disabled - pilus attachment is constant at $P_s = 0.5$. $\gamma = 0.5$, $k_U = 0.001$, ($1.35 \times 10^5s$,  1000s/frame). 
\end{itemize}

\subsection{Anisotropy parameter calculations}
For simulation data, anisotropy $\Delta S_{\phi}(r, d_{edge})$ was calculated as follows: for each cell, we calculated the entropy of the normalized histogram of the polar coordinates $\phi$ defining the vectors between the cell's centroid and the positions of $k$ neighbors within a radius $r$ defining the length scale of interest as $S_{\phi}(r) = \sum_{i}{p(\phi_{i}) \ln p(\phi_i)}$, where $i$ denotes one of $n$ discrete bins in the histogram, and $p(\phi_i)$ is the probability of randomly selecting a $\phi$ value within the $i$th bin. 

By comparing $S_{\phi}(r)$ to the value associated with the uniform distribution $S_{uni} = \ln n$ , we calculated an anisotropy parameter: $\Delta S_{\phi}(r) = [1 - S_{\phi}(r)/S_{uni}] - \varsigma(n, k)$ where $\varsigma(n, k) \approx \frac{n}{k}[0.37n^{-0.38} + 0.0045]$ is an empirically-derived correction factor necessary due to the non-minimum entropy of a finite sample from a continuous uniform distribution, valid for $k \gg n$. 

For Figs.~\ref{dS_exp_sim}(d,e,f) and \ref{dS_r_d}(a,b), the distance from the colony edge $d_{edge}$ was calculated as the shortest distance from the cell of interest and the convex hull of the configuration. 

For analysis of experimental data, the procedure was similar. However, the low-resolution images of {\it P. aeruginosa} used for comparison did not allow precise localization of cell centroids. To circumvent this limitation, the images [such as those shown in Fig.~\ref{FigES}(d-f)] were converted to binary using a standard thresholding procedure followed by manual cleaning of pixels not representing bacteria. These binary maps were then re-sampled to reflect the number of cells $\langle N \rangle$ expected to occupy the covered area. This estimation was achieved by calculating the average individual cell area $a_o$ from high-resolution images such as those represented in Fig.~\ref{Fig1}(a,b), and calculating $\langle N \rangle = A_{tot} / a_o$, where $A_{tot}$ is the total area covered by cells in the low-resolution binary image. Then, $\langle N \rangle$ filled pixels were randomly selected from the binary image, to approximate a true cell configuration. To avoid artifacts due to the square pixel grid, the position of each randomly selected pixel was randomized by adding a random perturbation $\Delta x \in [-w/2, ~w/2]$ where $w$ is the average cell width. The anisotropy of 20 such configurations were averaged for each of 4 replicate experiments at each gellan gum concentration investigated. These 4 replicates were then averaged to produce the anisotropy plots in Fig.~\ref{dS_exp_sim}(a-c). For these plots, $d_{edge}$ was calculated as the distance from the cell of interest and the portion of the convex hull representing the true colony edge (ignoring the borders of the imaging frame). To avoid edge artifacts in the anisotropy plots, only cells further than the length scale $r$ from the image frame were included in the analysis.


\subsection{Results of model variants lacking substratum deformation}

\begin{figure*}
\centering
{\includegraphics[width=\textwidth]{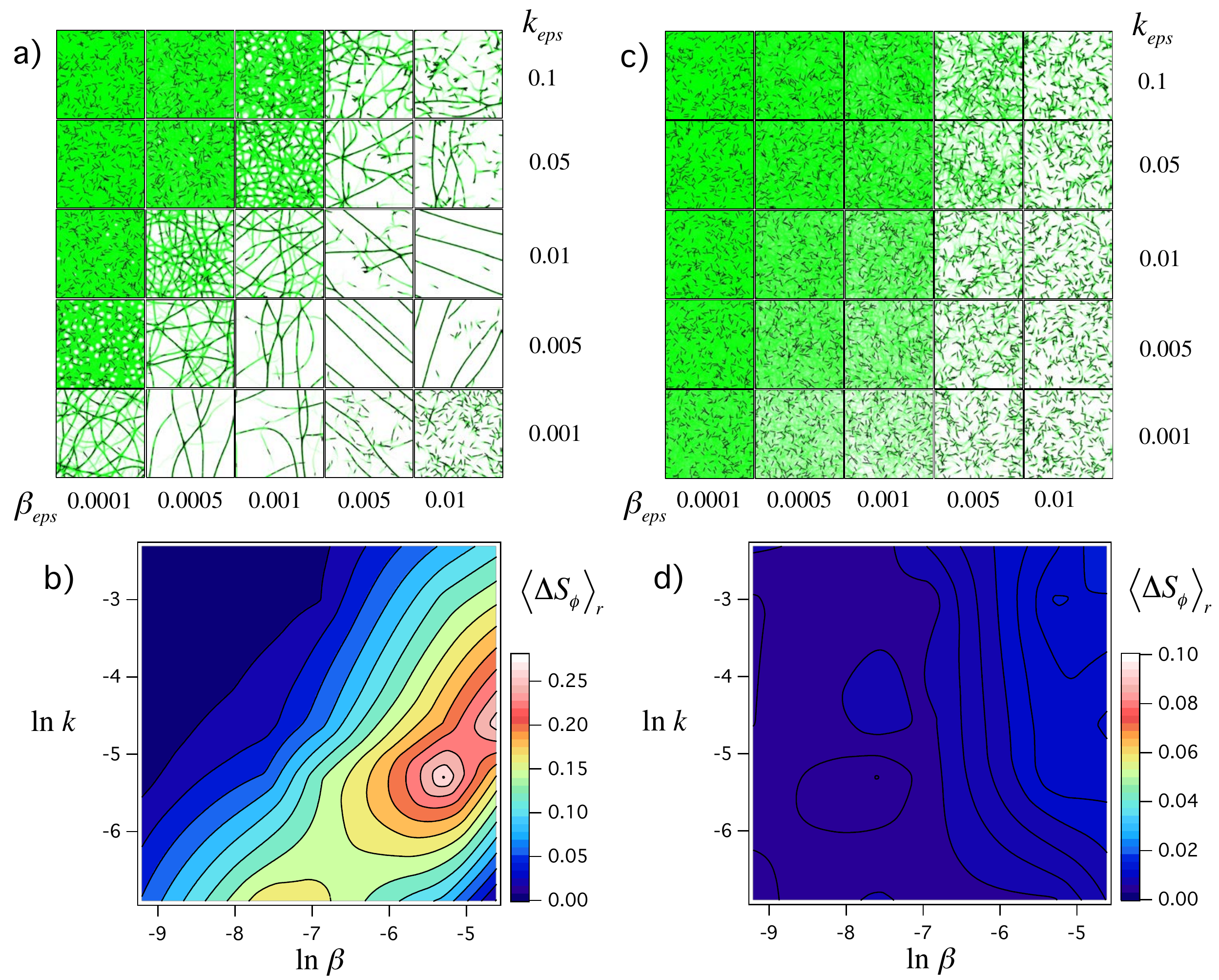}}
\caption{Results of model variants lacking substratum deformation and repulsive contacts between cells. The configurations in (a) and the corresponding anisotropy values shown in (b) were obtained by setting the pilus retraction period $t_{ret} = 0.1~s$, allowing cells to continuously sample EPS trails (illustrated in green). If trail sampling is stochastic ($t_{ret} = 10~s$), the resulting configurations (c) and anisotropy (d) show a complete lack of trail network formation.}
\label{EPS_no_rep}
\end{figure*}

To substantiate the claim that EPS trail following is not sufficient to produce the network patterns observed experimentally, we performed tests with several model variants that do not include the phenomenon of substratum deformation. The first of these is a highly idealized model that does not include cell-cell repulsion. That is, the simulated bacteria can occupy the same space. Results obtained with two subvariants of this model are provided in Fig.~\ref{EPS_no_rep}. In the first subvariant, we set the maximum pilus retraction period ($t_{ret} = 0.1~s$), 2 orders of magnitude faster than the experimentally derived period of $t_{ret}\approx 10~s$, used in the main text. The fast retraction period allows the bacteria to continuously sample the space in front of them, and follow EPS trails very effectively. Because cell-cell repulsion is disabled, the dynamics of the bacteria are determined primarily by the deposition rate ($k_{eps}$) and degradation rate ($\beta_{eps}$) of the EPS trails that modulate the pilus attachment probability. This is shown in Fig.~\ref{EPS_no_rep}(a), which illustrates steady-state configurations of fixed cell populations ($N = 250$) in a space with periodic boundary conditions, for varied $k_{eps}$ and $\beta_{eps}$ values. With continuous trail sensing and no repulsive interactions between cells, trail formation occurs over a large area of parameter space, as demonstrated by the high anisotropy values illustrated in the corresponding interpolated contour shown in Fig.~\ref{EPS_no_rep}(b) (here, $\langle \Delta S_{\phi}\rangle_{r}$ represents the spatial anisotropy averaged over length scales $r \in [10, 20, 30, 40, 50]w$ for the entire simulation duration of $t_f = 5 \times 10^4~s$, measured for each combination of $k_{eps}$ and $\beta_{eps}$). However, if the maximum pilus retraction period is increased to the experimentally constrained value of $t_{ret} = 10~s$, trail following does not occur for any of the values of $k_{eps}$ and $\beta_{eps}$ tested. This is demonstrated clearly by the resulting configurations shown in Fig.~\ref{EPS_no_rep}(c), and the corresponding low anisotropy values shown in Fig.~\ref{EPS_no_rep}(d).

So far we have demonstrated that network formation is extremely sensitive to the trail sampling rate. While we are confident in our estimation of the true sampling rate based on the available literature, it is possible that this value is sensitive to experimental conditions. Therefore, we continue our investigating of the continuous sampling condition in the second model variant by holding the retraction period at $t_{ret} = 0.1~s$ (continuous sampling), and enabling repulsive contacts between cells. Because physical contact between rod-shaped cells can lead to alignment of orientations and collective motion, its influence over the system's dynamics cannot be ignored. Because the frequency of cell-cell interactions can be expected to depend on density, we proceed by investigating two different cell populations. 

\begin{figure*}
\centering
{\includegraphics[width=\textwidth]{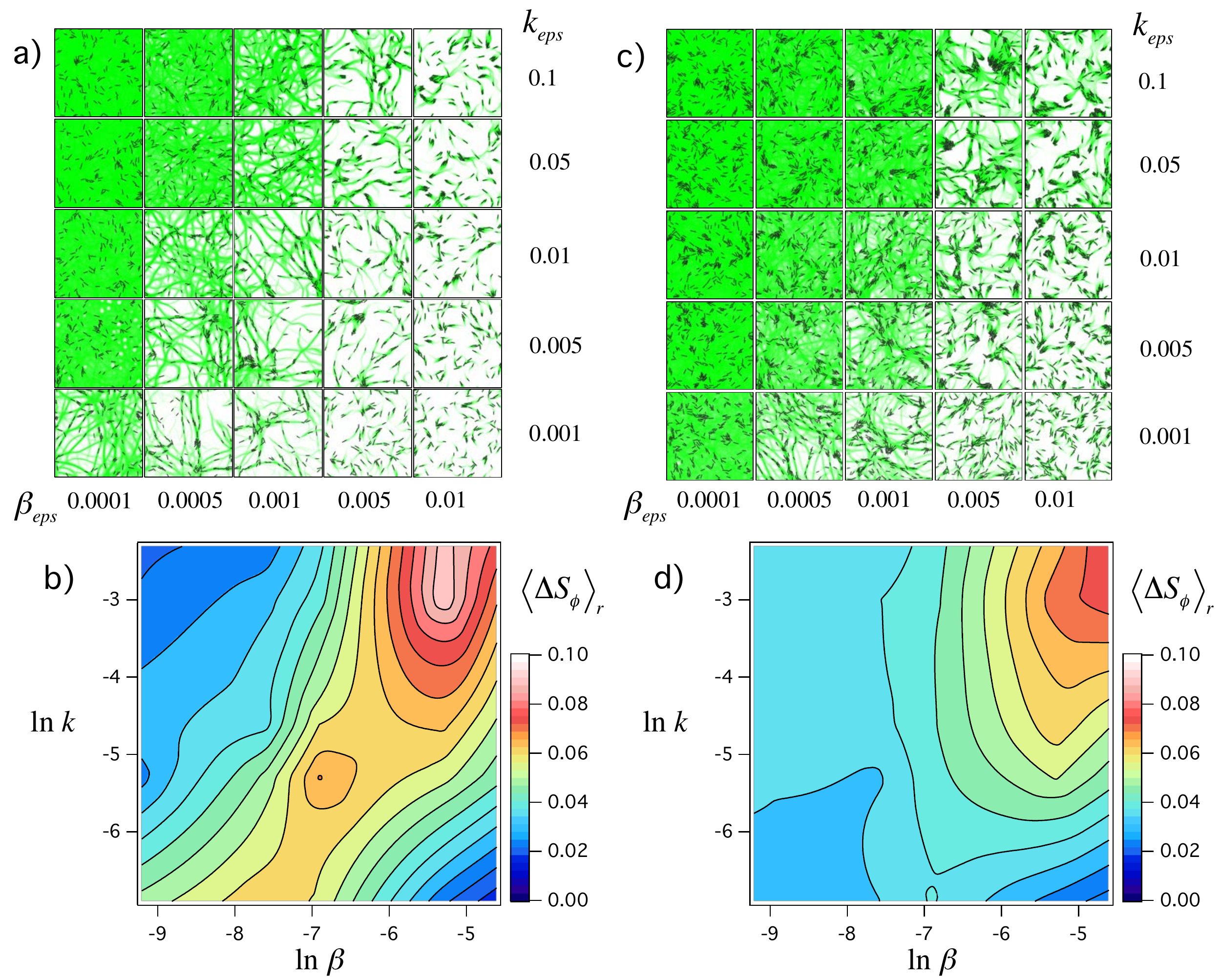}}
\caption{Results of model variants lacking substratum deformation, but including repulsive contacts between cells. Configurations of cells with EPS trails in green and the corresponding anisotropy values are shown for $N = 125$ (a, b), and $N = 250$ (c, d). In both cases represented, EPS trail sampling was continuous ($t_{ret} = 0.1~s$).}
\label{EPS_rep}
\end{figure*}

In Fig.~\ref{EPS_rep}(a), configurations of $N = 125$ cells are illustrated and, while network formation is not as pronounced as it is if repulsion between cells is ignored [Fig.~\ref{EPS_no_rep}(a)], it is still observable in configurations and the corresponding anisotropy values [Fig.~\ref{EPS_rep}(b)]. However, if the cell population is low and nutrients are plentiful, the population can be expected to increase. If the population is doubled to $N = 250$, network formation is no longer observed in configurations [Fig.~\ref{EPS_rep}(c)], and anisotropy is low for all values of $k_{eps}$ and $\beta_{eps}$ tested [Fig.~\ref{EPS_rep}(d)]. It is interesting to note that while trail network formation is not favored when repulsive contacts are included, the configurations in Fig. \ref{EPS_rep}(c) for high values of $\beta_{eps}$ and $k_{eps}$ appear to demonstrate nonequilibrium cluster formation and higher degrees of collective motion than would be expected without EPS trails. This apparent competition between collective motion effects and trail network formation deserves further study. 

To conclude this section, we have shown that trail network formation is not a robust consequence of EPS trial following due in part to the stochastic nature of spatial sampling with T4P. Even if that limitation is relaxed, physical interactions make the phenomenon very sensitive to cell density due to collective motion effects. For dilute populations of cells that can sample their environment continuously, EPS deposition may be enough for trail following phenomena. However, such behavior is not consistent with the biological system investigated in this work, and our model of emergent pattern formation must include substratum deformation to explain the experimentally observed phenomena.

\subsection{Results of model variants lacking EPS}

\begin{figure*}
\centering
{\includegraphics[width=\textwidth]{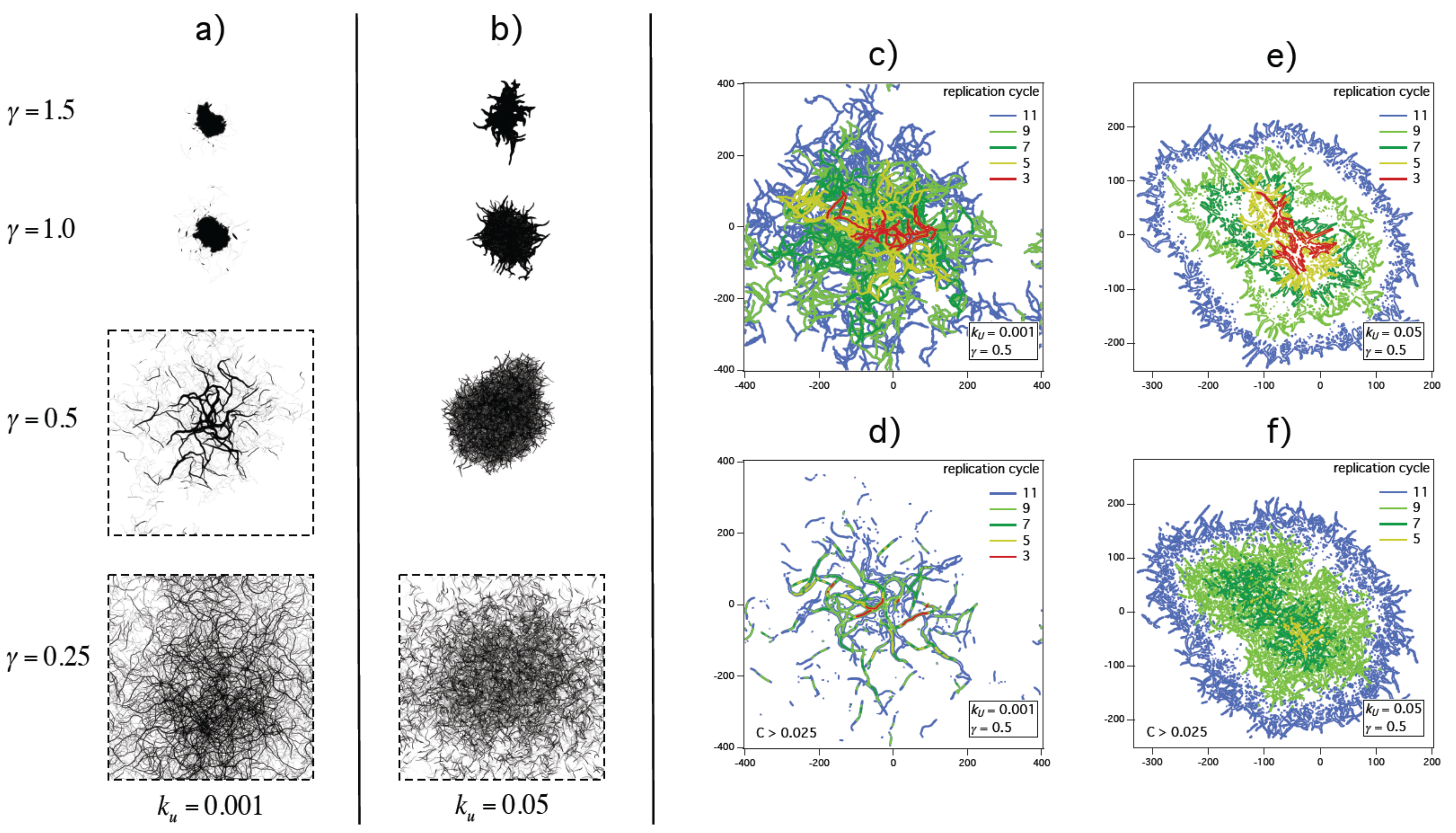}}
\caption{Colony morphologies simulated by the bacterial biofilm model without EPS deposition for constant surface attachment probability $P_s = 0.5$. Surface topography after 11 replication cycles shows dense, dendritic, network, and diffusive morphologies, depending on the substratum parameters $\gamma$ and $k_U$ (a, b). Dotted boxes indicate simulations where bacteria crossed the periodic boundaries before $t_f$. Contour plots (c-f) show areas of substratum deformation where $C_s >  0$ (c, e) and $C_s > 0.025$ (d, f) for the indicated replication cycles and substratum parameters. }
\label{SR2}
\end{figure*}

To detail the role of EPS in colony morphogenesis, we ran simulations using a model variant that lacks EPS deposition. Without EPS, pilus attachment probability $P_s$ is uniform throughout the simulation space. In this case (for fixed motility parameters as described in the model description in our earlier work \cite{zachreson2016emergent}) colony morphology depends only on $\gamma$, $k_U$, and $P_s$. We tested the model's behavior for $\gamma = 1.5, 1.0, 0.5, 0.25$, $k_U = 0.05, 0.001$, and $P_s = 0.5, 0.25$. Figure \ref{SR2} shows results for $P_s = 0.5$. A unique morphology was observed for $\gamma = 0.5$ and $k_U = 0.001$ [Fig.~\ref{SR2}(a) $\gamma = 0.5$]. Long, tapered trails with high cell density emerge near the colony interior, surrounded by a subpopulation of diffuse cells that move independently. Examination of the morphogenesis process (Supplemental Movie S8) reveals a pattern formation mechanism distinct from that involving the effects of EPS (Supplemental Movies S3, S4), as described in the main text.

This process occurs as follows: if the individual movement rate (as dictated by the attachment probability) is fast enough that the cells move before `sinking' into the substratum, they initially diffuse over the sufrace. Due to the stochastic nature of pilus binding and polarity reversal, the cells will at some point stop moving long enough to create a potential well in the substratum, trapping them in position. There, they slowly move back and forth, elongating the depression in the substratum while also growing and dividing. This process results in the observed elongated, tapered clusters that merge together to form the resulting dense network. The contour plots in Fig.~\ref{SR2}(c) illustrate how the tracks of the freely diffusing cells traverse the simulation space freely while Fig.~\ref{SR2}(d) illustrates how the network morphology coexists with the diffusing sub-population and remains confined to deep depressions in the substratum. 

If the deformation rate is too fast [Fig.~\ref{SR2}(b,e,f)], or if the attachment probability is lower (ie: when $P_s = 0.25$, not shown), cells sink too quickly to diffuse over the surface, favoring either an isotropic, dilute phase (if the substratum resistance is low) or a dense, dendritic phase if the substratum resistance is high enough to prevent individuals from escaping the expanding colony boundary.

Of course, even in the presence of EPS such morphologies could be expected if the affinity of pili to the unaltered surface was on the order of the constant attachment probability necessary for the formation of the patterns shown here. Therefore, EPS allows network formation to occur in conditions where the surface on which the bacteria are moving has a low affinity to T4P, and facilitates this process through a mechanism that is qualitatively distinct from that which occurs in its absence. 

\begin{figure*}
\centering
{\includegraphics[width=\textwidth]{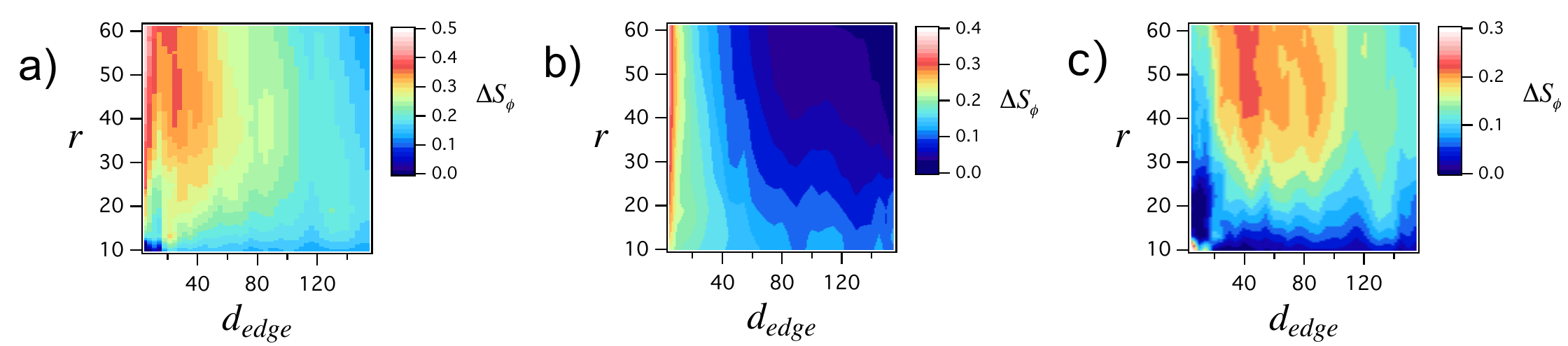}}
\caption{Intensity maps of $\Delta S_{\phi}$ as a function of length scale ($r$), and distance from the colony edge ($d_{edge}$) for simulated biofilms with substratum deformation only (a), EPS trail following as well as substratum deformation (b), and the corresponding residual (c), $c = |a - b|$. For both cases, substratum stiffness $\gamma = 0.5$ and deformation rate $k_U = 0.001$.}
\label{dS_r_d}
\end{figure*}

The trail network patterns that occur in the absence of EPS trail following are qualitatively and quantitatively distinct from those produced by the combination of EPS trail following and furrowing as presented in the main text. This is demonstrated in Fig.~\ref{dS_r_d} which shows a comparison of anisotropy maps for simulated colonies with the same substratum deformation rate $k_U = 0.001$ and stiffness coefficient $\gamma = 0.5$, without [Fig.~\ref{dS_r_d}(a)], and with [Fig.~\ref{dS_r_d}(b)] EPS trail following. The residual of these two plots [Fig.~\ref{dS_r_d}(c)] illustrates how the patterns formed due only to substratum deformation retain anisotropy over much longer length scales due to the long, tapered nature of the trails.

\subsection{Image processing and growth rate estimation}

The phase contrast microscopy time-series used to calculate growth rate was comprised of 1000 frames captured at 1 frame /2 s (2000 s). Phase contrast microscopy was performed using an Olympus IX71 inverted research microscope with a 100x 1.4 NA UPlanFLN objective, FViewII monochromatic camera and AnalySIS Research acquisition software (Olympus Australia, Notting Hill, VIC, Australia) fitted with an environmental chamber (Solent Scientific, Segensworth, UK).

The growth rates were determined by morphological analysis of the individual cells using a modified version of the BacFormatics image processing platform \cite{turnbull2016explosive} which was developed in MATLAB. The calculated growth rates were slightly faster than the actual ones due to single-pixel trimming during image segmentation. To correct for this, we set the individual growth rate to $k_g = 3.5 \times 10 ^{-4}$ $ \mu$m$s^{-1}$ in our simulations (Fig. 1c, dashed line). 

The source code for the beta version of the BacFormatics toolbox used for image segmentation and tracking in this work is available at: \\https://github.com/ithreeMIF/BacFormatics\_CZ/. 

\begin{figure*}
\centering
{\includegraphics[width=\textwidth]{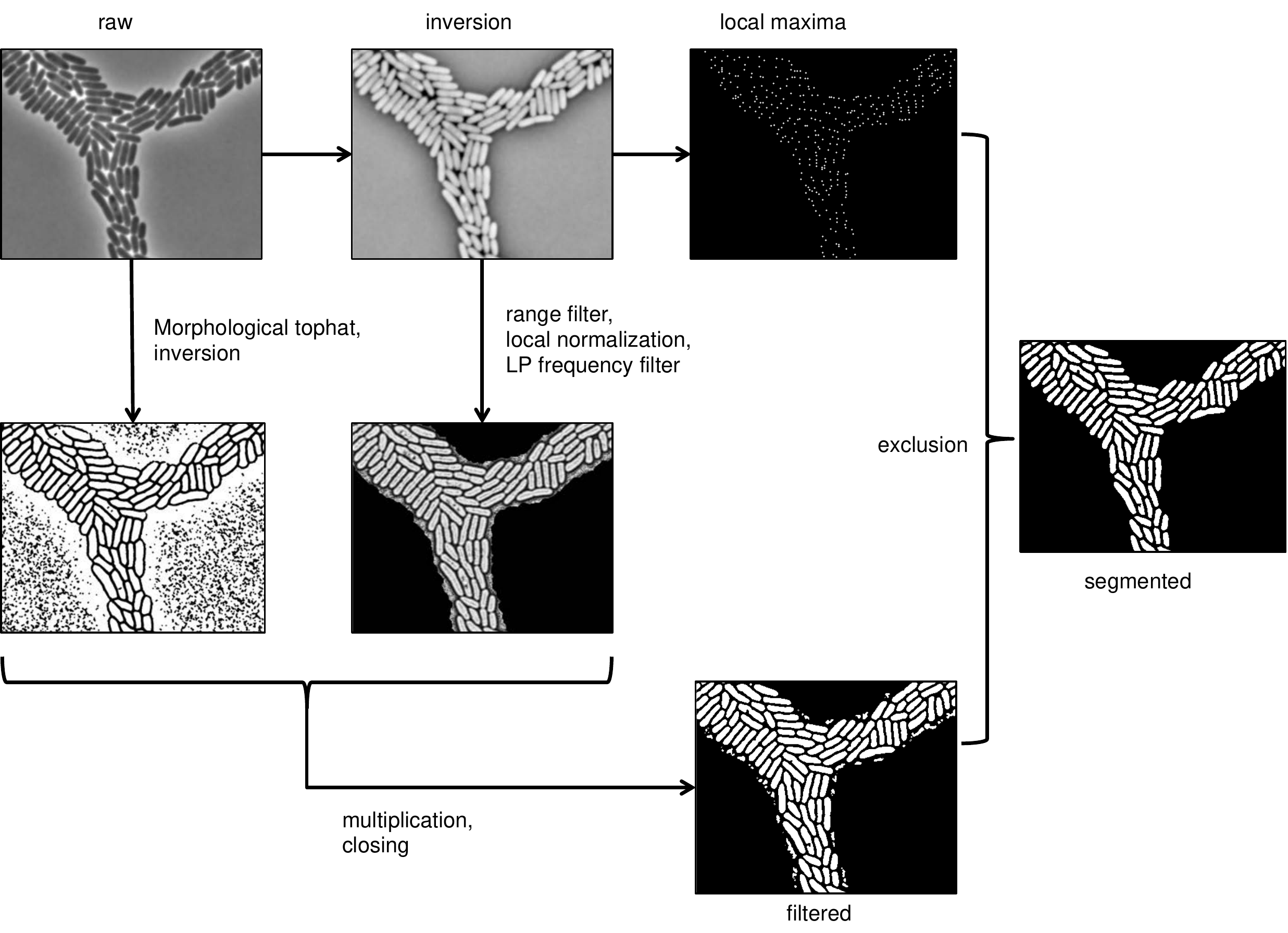}}
\caption{Schematic of the image segmentation algorithm used to analyze raw microscopy data. The raw phase-contrast images are translated into binary images containing only the pixels associated with bacterial cell bodies. The segmented images were used to estimate the growth rates of individual bacteria.}
\label{S1}
\end{figure*}

Briefly, the phase contrast images are segmented by applying a range filter, followed by local normalization \cite{locnorm}, low-pass frequency filtering (using a vectorized version of the butteroworth low-pass filter found here \cite{lpf}), and thresholding. The resulting binary image of the cell bodies is multiplied by another binary image that is created using morphological tophat filtering to isolate an image of the spaces between cells. The resulting binary image effectively segments individual cells. As a final step, segments are excluded from this image if they do not contain local maxima from an inverted version of the original phase-contrast image. This process is summarized schematically in Fig.~{\ref{S1}}.

Cell centroid positions, orientations, lengths $l$, and widths $w$ were estimated using the regionprops() function in the MATLAB image processing toolbox. Cell tracking was carried out by creating a distance matrix with the cell centroid positions in two consecutive frames and linking cell IDs based on the minimum distance traveled between frames.

Before estimating growth rate, we examined the length of each tracked cell as a function of time and identified discontinuities. These were identified based on a cutoff length variation between consecutive frames $\Delta l_{min} = 0.6$ $\mu$m (10 pixels) that is biologically unrealistic and is significant with respect to random variation in segment dimensions. Such discontinuities can occur naturally when cells divide, or when there are errors in segmentation that merge two or more cells, or split individual cells. 

Between such discontinuities, the aspect ratio $\kappa = l/\langle w \rangle_{t}$ was calculated for each frame where $\langle w \rangle_{t}$ is the time average of the cell width, which should not change as the cell elongates. We estimated the rate of change of the aspect ratio for each continuous timeseries of lengths by applying a linear regression to each data sequence exceeding a threshold duration ($\Delta l_{min} = 400s$, 200 frames). The slope of the linear regression accounts for an individual growth rate value, which we averaged to find the expectation value for growth rate in each experiment. The $k_g$ values in the main text (Fig. 1c) therefore represent linear approximations of the rate at which the aspect ratio of each cell is increasing. This is appropriate since our model uses the cell width as its fundamental unit of length $w = 1$ $\mu$m. 


\bibliography{prop_aniso_refs}

\end{document}